\documentclass[10pt]{article}
\usepackage{amsfonts}
\usepackage{mathrsfs}
\usepackage{amsmath}	
\usepackage{cite}
\usepackage{graphicx}
\usepackage{rotating}
\usepackage{hyperref}
\usepackage{verbatim}
\usepackage{slashed}
\usepackage[all]{xy}
\usepackage{xcolor}

\csname @addtoreset\endcsname{equation}{section}
\newcommand{\M}{{\cal M}}

\newcommand{\beq}{\begin{equation}}
\newcommand{\eeq}{\end{equation}}
\newcommand{\bea}{\begin{eqnarray}}
\newcommand{\eea}{\end{eqnarray}}
\newcommand{\ba}{\begin{array}}
\newcommand{\ea}{\end{array}}
\newcommand{\bit}{\begin{itemize}}
\newcommand{\eit}{\end{itemize}}
\newcommand{\nn}{\nonumber}

\newcommand{\mezzo}{\frac{1}{2}}
\newcommand{\complesso}{{\ \hbox{{\rm I}\kern-.6em\hbox{\bf C}}}}
\newcommand{\reale}{{\hbox{{\rm I}\kern-.2em\hbox{\rm R}}}}
\newcommand{\uno}{ \,  \raisebox{+0.14em}{{\hbox{{\rm \scriptsize ]}} \raisebox{-0.2em}{\kern-.8em\hbox{1}}}} \, }  

\newcommand{\p}{\partial}

\newcommand{\w}{\wedge}

\renewcommand{\a}{\alpha}
\renewcommand{\b}{\beta}

\renewcommand{\d}{\delta}
\newcommand{\D}{\Delta}
\newcommand{\e}{\epsilon}

\renewcommand{\k}{\kappa}
\renewcommand{\l}{\lambda}
\renewcommand{\L}{\Lambda}

\newcommand{\m}{\mu}

\newcommand{\n}{\nu}
\renewcommand{\r}{\rho}
\newcommand{\s}{\sigma}

\renewcommand{\S}{\Sigma}

\newcommand{\Om}{\Omega}

\begin{document}


\begin{titlepage}
\begin{flushright}
UAI-PHY-16/08
\end{flushright}
\vspace{2.5cm}
\begin{center}
\renewcommand{\thefootnote}{\fnsymbol{footnote}}
{\huge \bf Thermodynamics of Regular}
\vskip 5mm
{\huge \bf  Accelerating Black Holes}
\vskip 25mm
{\large {Marco Astorino\footnote{marco.astorino@gmail.com}}}\\
\renewcommand{\thefootnote}{\arabic{footnote}}
\setcounter{footnote}{0}
\vskip 10mm
{\small \textit{Departamento de Ciencias, Facultad de Artes Liberales,\\
  UAI Physics Center,
Univesidad Adolfo Iba\~{n}ez, \\ 
Av. Padre Hurtado 750, Vi\~{n}a del Mar, Chile}
}
\end{center}
\vspace{5.1 cm}
\begin{center}
{\bf Abstract}
\end{center}
{Using the covariant phase space formalism, we compute the conserved charges for a solution, describing an accelerating and electrically charged Reissner-Nordstrom black hole. The metric is regular provided that the acceleration is driven by an external electric field, in spite of the usual string of the standard C-metric. The Smarr formula and the first law of black hole thermodynamics are fulfilled. The resulting mass has the same form of the Christodoulou-Ruffini mass formula.\\
On the basis of these results, we can extrapolate the mass and thermodynamics of the rotating C-metric, which describes a Kerr-Newman-(A)dS black hole accelerated by a pulling string.
}
\end{titlepage}





\section{Introduction}
\label{intro}

The study of black hole thermodynamics has been a fruitful area of research since the pioneering discoveries of Bekenstein and Hawking \cite{bekenstein},\cite{hawking} about the black hole temperature, entropy and radiation. It provided a first insight into the relation between gravity and quantum mechanics. More specifically, some geometric features of the event horizon, such as its area or surface gravity, turn out to be linked to the thermodynamic properties of the black hole, such as its entropy or temperature. Given that thermal quantities of a physical system are related to the statistical description of its microstates, when we consider black hole thermodynamics, we are addressing its underlying microscopic degrees of freedom possibly associated to some quantum features of gravity \cite{novikov}. \\
The thermodynamics of several kinds of black holes have been studied in the last decades in general relativity and other gravitational theories. But only recently some attempts in the study of thermodynamics of accelerating black holes have been made \cite{ray-boost} - \cite{ruth}. There are basically two reasons for this.
First of all, the accelerating black holes are described by C-metrics \cite{Griffiths}, \cite{griffiths-interptering}, which have an uncommon asymptotic behaviour: there is not a constant curvature background at infinity, and moreover, depending on the range of the parameters we are considering, an accelerating horizon may also occur to complicate the asymptotic structure. In particular, the unconventional falloff of the metric at large radial distances makes it difficult to identify the timelike Killing vector necessary to compute the conserved charge associated with the mass of the black hole. Since the mass value enters in the laws of black hole thermodynamics, it is fundamental to have some good criteria to select the normalisation of the above Killing vector. \\
The second reason, which makes the study of the accelerating black hole thermodynamics unclear, is that these C-metrics present at least one non-removable conical singularity on the azimuthal axis of symmetry, both in the static and in the rotating case. It is not clear to the author if the first law of black hole thermodynamics still has to hold in case of singular black hole metrics\footnote{After the pubblication of this paper a treatment about thermodynamics in presence of conical defect appeared \cite{ruth-cony}}.\\
Therefore, in this paper we pursue a slightly different approach based on a variation, originally found by Ernst \cite{ernst-remove}, of the accelerating and magnetically charged black hole, which consists in embedding the accelerating Reissner-Nordstrom (RN) solution in an external magnetic field. On the other hand, the metric we will specifically study is immersed in an external electric field, which is responsible for the black hole acceleration instead of the usual singular cosmic string (or strut). The presence of the external electric or magnetic field makes it possible to remove all the conical singularities \cite{ernst-remove}. Thus the metric we will focus on in this paper is fully regular outside the event horizon.\\
The choice of working with a regular solution is significant because it directly removes the second difficulty mentioned above, about the study of accelerating black hole thermodynamics.  \\
Moreover, the regularity of the solution is an essential ingredient in the covariant phase space methods \cite{wald},\cite{Barnich-Stokes},\cite{Barnich-Compere} and  \cite{gao} to compute the conserved charges. These techniques are particularly useful in the case of unconventional asymptotic behaviour because they allow one to compute conserved charges in a finite region surrounding the black hole, thus bypassing the criticalities placed at very large distances from the black hole.  \\
This approach already turned out to be successful in the study of similar spacetimes with non trivial asymptotic behaviour, as shown in  \cite{mass-mkn}. It has been confirmed, in the extremal case, with different methods, as the near horizon analysis \cite{mkn-cft}. We would also like to extend these results for accelerating black holes; this will be done in Section \ref{mass}, while in Section \ref{acc-ectr} we introduce the accelerating black hole immersed in an external electric field background, a sort of dual of the Melvin magnetic universe. \\
Unfortunately, in the presence of the cosmological constant, the solution generating methods to build the electromagnetic background surrounding the black hole is not available; thus we cannot remove the conical singularity from the accelerating solution in this way. In that case the metric remains irregular, so we cannot proceed with the covariant phase space formalism to compute conserved charges. At most, extrapolating the results of Section \ref{mass}, we can present, in Appendix \ref{app1}, some proposal to generalise the mass and the thermodynamics of the rotating C-metric in the presence of the cosmological constant. \\

\section{Accelerating charged black hole in an external electric field}
\label{acc-ectr}

The equations of motion for Einstein-Maxwell theory, governed by the action 
\beq
     I[g_{\m\n},A_\m] = - \frac{1}{16\pi}  \int_M d^4x  \sqrt{-g} \left( R -  F_{\m\n} F^{\m\n} \right)  \quad ,
\eeq 
are
\bea \label{eom}
       & &   R_{\m\n}  -  \frac{R}{2} \  g_{\m\n} \hspace{2mm} =  \ 2  \left( F_{\m\r}F_\n^{\ \r} - \frac{1}{4} g_{\m\n} F_{\r\s} F^{\r\s} \right)  \quad ,     \\
  \label{eom2}     & &  \partial_\m \left( \sqrt{-g} F^{\m\n} \right) \ = \ 0 \qquad .
\eea
We consider the following solution for these equations:
\bea  
      \label{metric}  ds^2    &=&   \frac{ [\L(r,x)]^2 }{(1 + A r x )^2} \left[ - \frac{G(r)}{r^2} dt^2 + \frac{r^2 dr^2}{G(r)} + \frac{r^2 dx^2}{H(x)} \right] + \frac{r^2 H(x) \D_\varphi^2 \ d\varphi^2 }{(1+A r x)^2 [\L(r,x)]^2 }    \quad , \\
      \label{Apot}    A_\m  &=&   \big[ A_{t_0} + A_t(r,x)  , 0,0,0 \big]      \quad , 
\eea
where
\bea
      G(r)     &:=& (1-A^2 r^2) (r-r_+)(r-r_-)  \quad ,   \\
      H(x)     &:=& (1-x^2) (1 + A r_+ x)(1+Ar_- x)   \quad ,   \\
      \L(r,x)  &:=&  \left (1 +  \frac{q E x}{2} \right)^2  +  \frac{ E^2 r^2 H(x)}{ 4 ( 1 + A r x )^2}  \quad ,  \label{2.9} 
      \eea
$$    A_t(r,x) :=  \frac{ 2 A E r ( 2 m r - q^2 ) - 4 A^2 q r^2 + E^2 q \left[ (2 m - r) r - q^2 (1 - A^2 r^2)  \right]}{4 A^2 r^3} + \frac{E G(r) \left[ 2 A r + q E (1 + 2 A r x) \right]}{4 A^2 r^3 (1+Arx)^2} \ . $$

\begin{figure}[h!]
  \centering
 \includegraphics[scale=0.7]{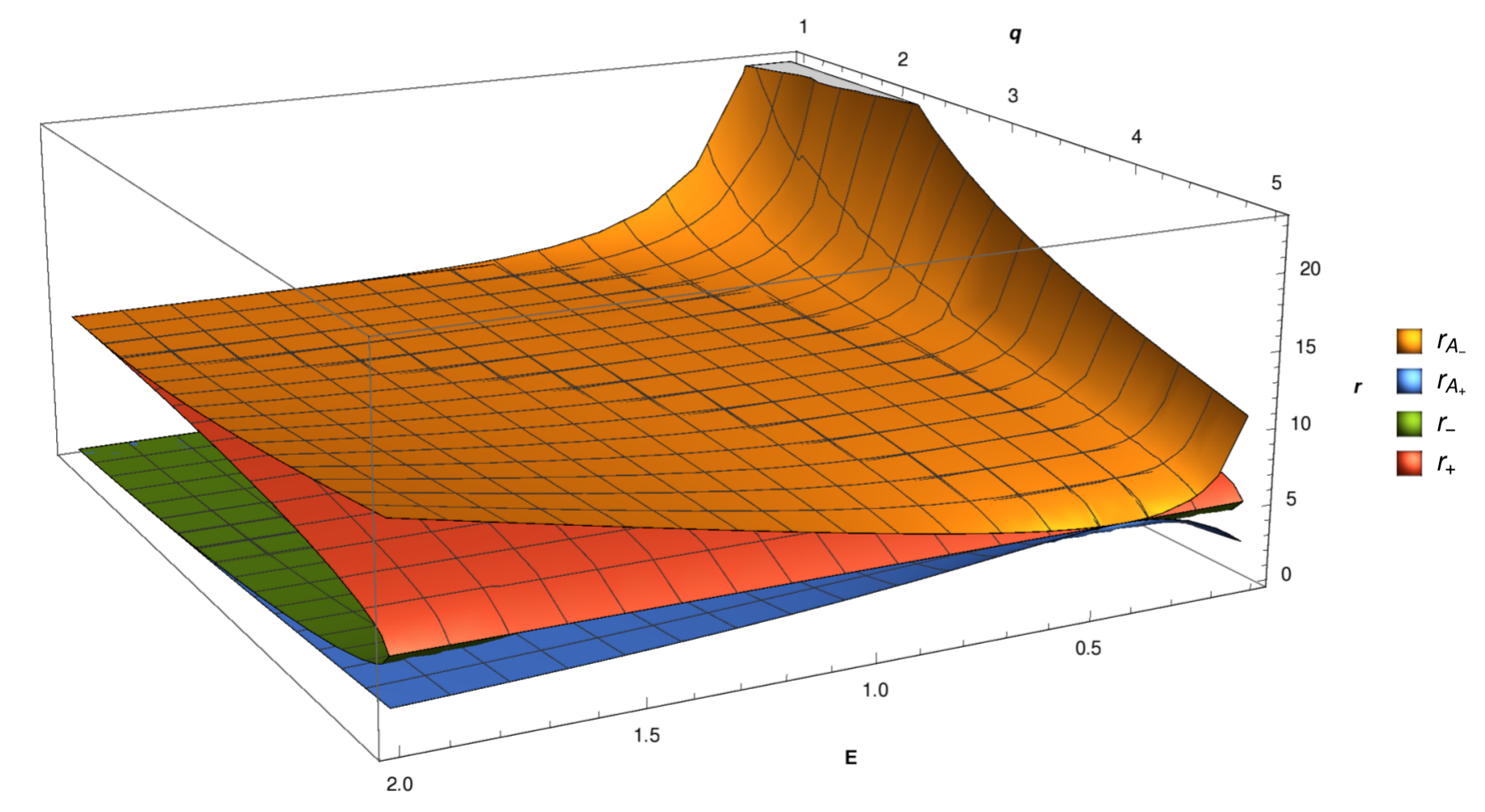} 
  \caption{\small Positions of the Killing horizons for a fixed value of the mass parameter ($m=5$). The picture does not change qualitatively for different values of m or range of $E$ and $q$.}
\label{raggi} 
 \end{figure}

\begin{figure}[h!]
  \centering
 \includegraphics[scale=0.7]{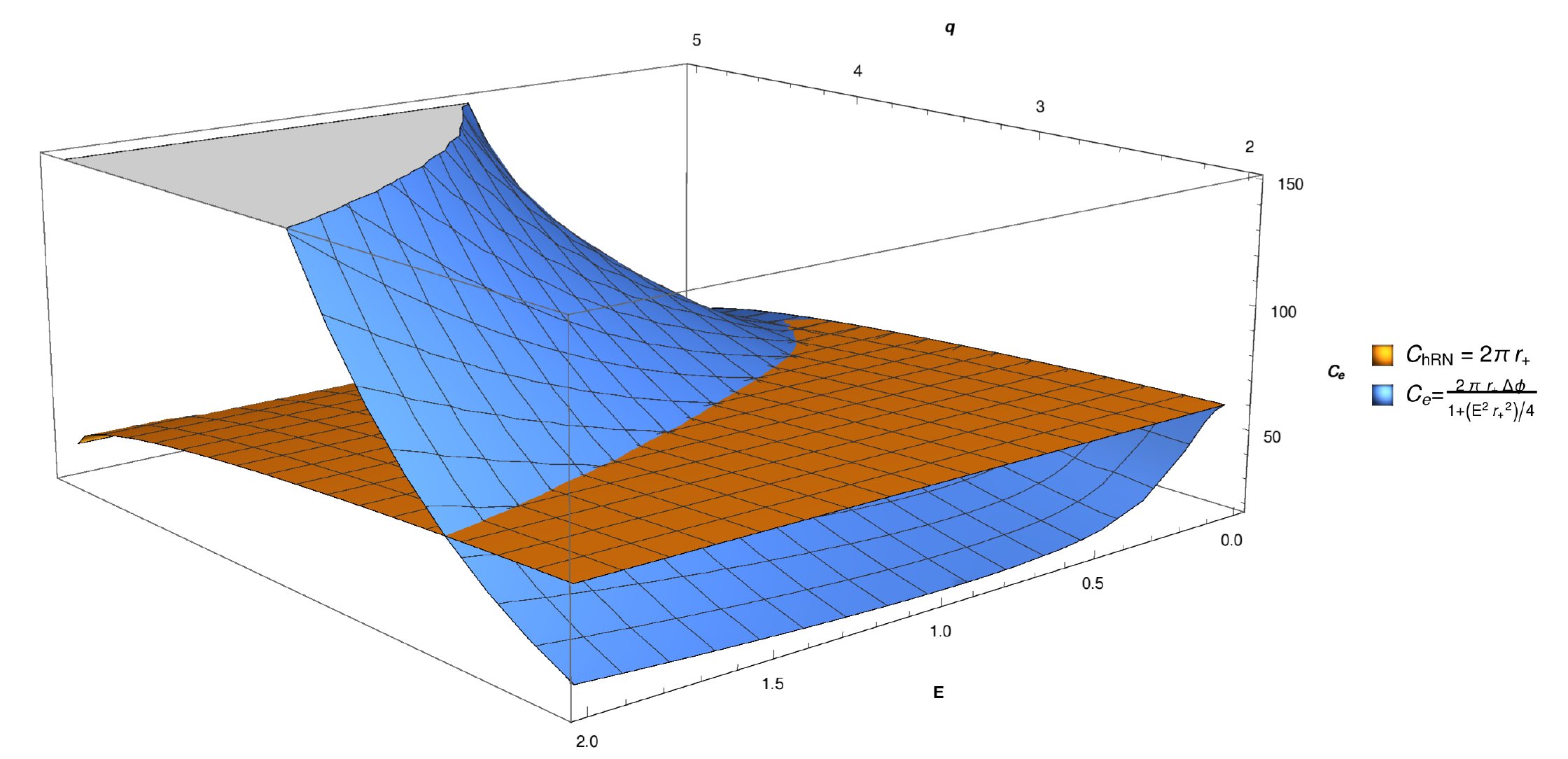} 
  \caption{ \small Plot of the equatorial circumference of the RN event horizon $C_{hRN} = 2 \pi r_+$ versus the polar circumference $C_e$, for $q\in[2,5]$ , $E\in[0,2]$ and for a fixed value of the mass parameter $m=5$. Depending on particular parametric values, the electrified equatorial circumference can either shrink or expand with respect to the standard RN.}
\label{figu1} 
 \end{figure}

The electromagnetic gauge potential $A_\m$ is related to the electromagnetic field strength tensor, as usual,  by $F_{\m\n}=\p_\m A_\n -\p_\n A_\m$. \\ 
The solution (\ref{metric})-(\ref{2.9}) have axial symmetry, therefore both the metric $g_{\m\n}$ and the gauge electromagnetic potential $A_\m$  depend explicitly on the $(r,x)$ coordinates only. The physical parameters involved in the solution are $ A, E, m$ and $ q $, respectively related to the acceleration, the intensity of the external electric field, the mass and the intrinsic electric monopole charge of the black hole. We will consider, without loss of generality the positive A branch, the $A<0$ branch is specular (and it is equivalent to inverting the range of the polar angle). The interpretation of the parameter $A$ as the acceleration can be inferred by studying the weak field limit of the above metric, i.e. when $m<<A$ \cite{griffiths-interptering} -\cite{acc-btz}. The radial coordinate $r$ is, of course, positive and $x \in [-1,1] $ since it corresponds to the polar angle (of spherical coordinates) through the relation $x=\cos \theta$.   \\    
These kinds of solutions were firstly generated by Ernst in \cite{ernst-remove} by applying a Harrison transformation to the charged C-metric. We recall that C-metrics describe a pair of casually disconnected black holes which accelerate away from each other along opposite directions under the action of forces due to the presence of cosmic strings or struts represented by conical singularities. When the nodal singularity is a deficit angle, the $\d$-like associated contribution to the energy momentum tensor is interpreted as a pulling string, while the excess angle is interpreted as a pushing strut, for a recent review about C-metrics see \cite{griffiths-interptering}, \cite{Griffiths}. 
Ernst has found out that it is possible to remove the conical singularity from the charged and accelerating black hole, described by the C-metric, because the Harrison transformation introduces a background external magnetic field\footnote{The Harrison transformation is a one-parameter element of the symmetry group of the solution space of axisymmetric and stationary spacetimes in Einsten General Relativity (without cosmological constant) coupled with Maxwell electromagnetism, $SU(2,1)$. The parameter, the Harrison transformation is introducing, is reprented by $E$ and it is directly related to the intensity of the external electromagnetic field. Hence, when $E=0$ the Harrison transformation reduces to the identity element and the seed solution remains unchanged, no additional external electromagnetic field is added.}. Therefore there is no need for a cosmic string or strut that provides the acceleration to the black hole. In the Ernst solution the acceleration is furnished by the interaction between the monopole charge of the black hole and the external electromagnetic field. In order to remove the conical singularity, the external and intrinsic electromagnetic field of the solution have to be of the same kind. Otherwise the extra conical singularity brought in by the Lorentz-like interaction between the intrinsic black hole charge and the electromagnetic background is not sufficient to compensate the C-metric axial deficit (or excess) angle, as shown in \cite{rot-pair}. For this reason, originally, a magnetically charged accelerating RN black hole was considered in an external magnetic field. In the case considered by Ernst, for small values of the acceleration, the metric approaches the Melvin magnetic universe \cite{melvin}.\\
On the other hand, in this paper we prefer to consider the electromagnetic dual\footnote{In 4D the electromagnetic duality allows one to interchange the role of the electric and magnetic fields. In particular a given solution of the Einstein-Maxwell equations \ref{eom}-\ref{eom2} ($g_{\m\n},F_{\m\n}$) can be mapped in another solution ($g_{\m\n},*F_{\m\n}$) of the same equations of motion, where the Hodge dual of the  electromagnetic field $F_{\m\n}$ is given by $*F_{\m\n} = \mezzo  \epsilon_{\m\n\s\r} F^{\s\r}$} of the original Ernst solution because its electromagnetic potential $A_\m$ is better behaved on the axial axes. In fact the dual Ernst solution (\ref{metric})-(\ref{2.9}) has no problems related to Dirac strings due to the presence of the intrinsic magnetic monopole. Of course, the dual Ernst solution asymptotically approaches the Melvin magnetic universe dual, which is a cylindrical-symmetric electric geon, with intensity proportional to the electric field parameter $E$. Note that, since the presence of the external electromagnetic field, codified by the function $\L(r,x)$ in the Ernst metric (and its dual) is factorized out from the $(t,r,x)$ part of the metric, its Penrose diagram and its conformal structure remain the same as the standard charged C-metric. \\
The good behaviour of the dual Ernst metric makes it suitable for analysing its conserved charges with the covariant phase space methods \cite{wald} -\cite{Barnich-Compere}, which have recently been revealed to be quite convenient in the case of non-constant curvature asymptotics \cite{barnich-goedel}, \cite{mass-mkn}. That solution have been useful, for instance in \cite{hawking-ross}-\cite{brown}, to clarify some issues related to the electric-magnetic duality at the (semi-classical) quantum level. \\
The constant $\D_\varphi$ can be set to ensure the absence of a conical singularity  at one of the two poles. A four-dimensional spacetime contains conical (or nodal) singularities if the two-dimensional base manifold metric $d\hat{s}^2$, obtained from the full metric fixing $t$ and $r$ to  some constant, cannot be cast, around the poles, in the form $d\hat{s}^2 = d\theta^2 + \sin^2 \theta d\varphi^2 $, when using spherical coordinates. While in $(x=\cos \theta,r)$ coordinates this means that the two-dimensional surface metric cannot be cast in the form
$$ d\hat{s} = \frac{dx^2}{1-x^2} + \D_\varphi^2 (1-x^2) d\varphi^2   $$
with $\D_\varphi=1$. Thus the surface has a conical singularity at the poles, with a deficit (or excess) in the azimuthal angle given by $2\pi\d$, where $\d=1-\D_\varphi$. Hence for $\D_\varphi=1$ there are no nodal singularities.  In practice considering a small circle around the $x= \pm 1$ axis we can require regularity with the following constraint:
\beq       \label{con-sin}
      \frac{\textrm{circumference}}{\textrm{radius}} = \lim_{x \rightarrow \pm 1} \frac{2 \pi}{1-x^2} \sqrt{\frac{g_{\varphi \varphi}}{g_{xx} }} =  \frac{32 \pi \D_\varphi (1 \pm 2 A m + A^2 q^2)  }{(2 \pm q E)^4} = 2 \pi \quad .
\eeq   
Because of the asymmetry of the deficits (or excesses, depending on the values of the parameters) angle at the poles of the two hemispheres, just one of the two nodal singularities can be simultaneously removed by fixing $\D_\varphi$. We choose, without loss of generality, to regularise the north pole one ($x=+1$) by setting $\D_\varphi$ as follows:
\beq \label{acc-con}
        \bar{\D}_\varphi = \frac{(2 + q E)^4}{16 (1 + 2 A m + A^2 q^2)} \quad .
\eeq  
Note that, as there is no dependence on the radial coordinate, eq. (\ref{acc-con}) removes the conical singularity from the whole semi-axis defined by $x=1$.
The second nodal singularity, located on the southern semi-axis identified by $x=-1$, can be removed thanks to the presence of the external electric field, imposing
\beq\label{acc-con2}
      \frac{16(1-2Am+m^2A^2) \bar{\D}_\varphi}{(2-qE)^4} =1
\eeq
Therefore the acceleration parameter $A$ is constrained in terms of the other parameters of the metric $m,q,E$ 
\beq \label{Abar}
       \bar{A}_{\pm}(m,q,E) = \frac{m (16 + 24 E^2 q^2 + E^4 q^4 ) \pm \sqrt{m^2 \left( 16 + 24 E^2 q^2 + E^4 q^4 \right)^2- 64 q^4 E^2 \left(4 + E^2 q^2\right)^2}}{8 q^3 E \left(4 + E^2 q^2 \right)}
\eeq
to ensure full regularity of the solution outside the event horizon\footnote{No scalar invariants are divergent  nor are conical singularities present outside the event horizon.}. Hence, from a physical point of view, the acceleration is furnished by the interaction between the electric charge of the black hole and the external electric field, making the singular cosmic string unnecessary.\\
From eqs. (\ref{acc-con2}) and (\ref{acc-con}), in the weak field limit (for small values of the acceleration parameter $A$ and electric field parameter $E$) the classical Newtonian force of a charged particle in a uniform electric field can be recovered
\beq
            mA=qE \ \ .
\eeq
The constant $A_{t_0}$ is arbitrary, so it can be set such that the gauge potential has a well-defined vanishing acceleration limit. It can be done by an ad-hoc prescription that cancels the divergent term for $A \rightarrow 0$, i.e. requiring that the lowest order of the Laurent series in the $A$ parameter is $ - \frac{E}{2A} $, but of course in this way $A_{t_0}$ is not uniquely fixed. Indeed all the positive A-powers in the series expansion remain basically unconstrained. \\
Otherwise  $A_{t_0}$ can be uniquely fixed by the integrability condition of the mass, selecting the canonical frame, where the Coulomb electro-chemical potentials $\Phi_{int}$ is null\footnote{Although it is sufficient to require a less restrictive property, such that the electric potential is not singular.} on the surface of integration. Remarkably, in that case, the term that avoids divergences naturally appears. More details are elucidate in the next section. However in both cases we smoothly recover or the Reissner-Nordstrom black hole embedded in an external electric field for $A \rightarrow 0$ or the standard charged C-metric for $B \rightarrow 0$. \\
The black hole described by the metric (\ref{metric}) has the same conformal structure of the accelerating Reissner-Nordstrom black hole, because the contribution of the external electric field to the conformal factor in front of the $(t,r)$ part of the metric is not singular, within the accelerating horizon. In fact the black hole has an inner and an outer horizon $r_{\pm}$, which coincide with the RN ones, and an accelerating horizon $r_A$ respectively located where $G(r)=0$ :
\beq
        r_{\pm} = m \pm \sqrt{m^2-q^2}  \hspace{1.2cm} , \hspace{1.2cm} r_{A} = \frac{1}{A} \quad .
\eeq
The regularity constraint (\ref{Abar}) can take two values, then the accelerating horizon can be located, as shown in Figure \ref{raggi}, with respect to the inner and outer horizons as follows
\beq
       r_{A_+} <  r_- < r_+ < r_{A_-} \quad .
\eeq
Therefore two different pictures are possible from (\ref{Abar}).  Taking $\bar{A}_+$ the position of the accelerating horizon remains inside the inner horizon of the black hole. On the other hand, considering $\bar{A}_-$ as regularity constraint, the accelerating horizon is located, more conventionally, outside the black hole event horizon. It can be shown that for $A=\bar{A}_-$ some quantities, such as the  electric charge, have better defined limits in the case of a vanishing external magnetic field. This case can also be considered the standard branch because in the vanishing electric field limit, $E \rightarrow 0$, the value of the constrained acceleration $\bar{A}_-$ goes to zero, thus the regular RN black hole is retrieved. Instead, the acceleration regularity constraint  (\ref{acc-con}) is divergent for $\bar{A}_+$, in the null electric field limit. Henceforward, for these reasons, we will consider the value $\bar{A}_-$ as the constrained acceleration and we will omit, for ease of notation, the $-$ sign. \\
The external electric field strongly modifies the geometry of the event horizon, which is stretched or squashed in the axial direction, depending on the values of the free physical parameters $m, q, E$.  This can be inferred by inspecting the equatorial circumference of the event horizon embedded in the external electric field 
\beq
      C_e  = \int_0^{2\pi} \sqrt{g_{\varphi\varphi}} \ d\varphi = \frac{2 \pi r_+ \bar{\D}_\varphi}{1 + \frac{E^2}{4} r_+^2} \ \ .
\eeq
In fact, comparing $C_e$ with the standard RN equatorial circumference $2 \pi r_+$, we can appreciate, in Figure \ref{figu1}, how the variation of the parameters $p$ and $E$ (while $m$ is considered fixed in the plot) affects the size of the equatorial circumference. Thus, in the presence of acceleration, the deformation of the black hole event horizon is not just axially stretched, as it occurs in the pure Melvin background \cite{wild}. \\

\section{Mass, Electric Charge and Thermodynamics}
\label{mass}

The area of the event horizon is modified, with respect to the accelerating RN black hole, by the external magnetic field only through the factor $\D_\varphi$
\beq \label{area}
           \mathcal{A} = \int_0^{2\pi} d\varphi \int_{-1}^1 dx \sqrt{g_{\varphi \varphi} g_{xx}} \ = \   \frac{ 4 \pi \D_\varphi \ r_+^2}{1-A^2 r_+^2} \quad .
\eeq
The intrinsic electric and magnetic monopole charges can be computed, for the regularised black hole. They are respectively
\bea \label{Q}
      \mathcal{Q} &=& \frac{1}{8\pi} \int_\S F^{\m\n} d\S_{\m\n} = - \frac{1}{4\pi} \int_0^{2\pi} d\tilde{\varphi}  \int_{-1}^1 dx \ \sqrt{g_{\small \S}} \ n_\m \s_\n F^{\m\n}  = \frac{4 \ q \ \bar{\D}_\varphi}{4 - q^2 E^2} \quad ,\\
      \mathcal{P} &=& \frac{1}{4\pi} \int_\S F_{\m\n} \ dx^\m \wedge dx^\n =  0   \quad ,  
\eea
where $ d\S_{\a\b} = - 2 n_{[\a}\s_{\b]} \sqrt{g_S} \ d\varphi dx $ and $ \sqrt{g_{\small \S}} = \sqrt{g_{xx} g_{\tilde{\varphi} \varphi}} $ defines the two-dimensional volume element of the surface $\S$ which surrounds the horizon for fixed time and radial distance. $n_\m$ and $\s_\n$ identify two mutual orthonormal vectors, respectively time-like and space-like, which are normal also to the integrating surface $\S$.\\
The Hawking temperature, in the units we are considering where $c=\hbar=k_B$, is defined as usual in terms of the surface gravity $\k_s$
\beq \label{T}
            T = \frac{\k_s}{2\pi} \ \ .
\eeq
Considering\footnote{This vector is defined up to a constant rescaling. For the moment we naively take the simple normalisation characteristic of asymptotically flat spacetimes, but we will see that, since the non-trivial fields fall-off, a proper rescaling will be necessary.} $\chi=\p_t$ as the Killing vector generating event horizon, the surface gravity on that Killing horizon is given by
\beq \label{ks}
       \k_s = \sqrt{-\frac{1}{2} \nabla_\m \chi_\n \nabla^\m \chi^\n } \ \ \Bigg|_{r_+} = \frac{r_+ -r_-}{2 \ r_+^2} \ (1-A^2 r_+^2) \ \ .
\eeq
The Coulomb electric potential on the event horizon is 
\beq \label{Phi}
        \Phi_{r_+} := - \chi^\m A_\m \ \bigg|_{r=r_+} =  - A_{t_0} - \frac{E}{2 A} + \frac{q}{r_+} \left( 1- \frac{q^2 E^2}{4} \right) \ \ .
\eeq
Note that, also for this quantity, the well behaviour of the vanishing accelerating limit requires that the lowest order term in the Laurent expansion of $A_{t_0}$ coincides with $- \frac{E}{2 A}$.\\
The calculation of the mass is not that direct, basically because it is not clear which is the normalisation of the canonical generator associated with the conserved energy. In fact the involved asymptotic structure of the solution makes it hard to identify a global timelike Killing vector at infinity. We will proceed in an operational way, based on the covariant phase space methods and integrability \cite{wald} - \cite{gao}, which was successfully pointed out in \cite{mass-mkn}, for non accelerating black holes. We will compute the infinitesimal energy change due to the variation of the independent solution parameters ($\d m , \d q , \d E  $)\footnote{Of course the acceleration parameter $A$ can be equivalently considered instead of one between $m, q$ or $E$, inverting the constraint (\ref{Abar}). We have pursue this choice only for algebraic economy.}.\\
Generically the infinitesimal conserved surface charge associated with the the Killing symmetry $(\xi,\l)$ is given by
\beq \label{Qt}
      \d \mathcal{Q}_{(\xi,\l)} =  \oint_\S \d k_{(\xi,\l)} \left[ \d g_{\m\n} , \d A_\m ; g_{\m\n} , A_\m \right] \quad .
\eeq 
The conserved super-potential associated with the generic symmetry parameter $(\xi,\l)$ is given, for  Einstein-Maxwell theory, by 
\beq
      \d  k_{(\xi,\l)} \left[ \d g_{\m\n} , \d A_\m ; g_{\m\n} , A_\m \right] =  - \d K_{(\xi,\l)} + K_{(\d \xi, \d \l)} - \iota_\xi \Theta \quad ,
\eeq  
where
\bea
       K_{(\xi,\l)} [g_{\m\n},A_\m] &=&  \frac{\sqrt{-g}}{16 \pi G} \left( d x^{n-2}\right)_{\m\n} \ \left[ \nabla^\m \xi^\n - \nabla^\n \xi^\m  + 4 F^{\m\n} (\xi^\s A_\s + \l) \right]              \quad ,   \\
       \Theta \left[ \d g_{\m\n} , \d A_\m ; g_{\m\n} , A_\m \right]     &=&    \frac{\sqrt{-g}}{16 \pi G} \left( d x^{n-1}\right)_{\m} \left[ \nabla_\s \d g^{\m\n}  - \nabla^{\m} g^\n_{\ \n} + 4 F^{\s\m} \d A_\s \right] \quad ,
\eea
and
\beq
      \left(  d x^{n-p} \right)_{\m_1 ... \m_{n-p}}  = \frac{1}{p! (n-p)!} \e_{\m_1...\m_n} dx^{\m_{n - p + 1}} \w ...\w dx^{\m_n} \quad .
\eeq
Here $ \iota_\xi := \xi^\m \frac{\p}{\p dx^\m} $ defines the interior product between differential forms. \\ 
In this formalism the electric charge (\ref{Q}) associated to the symmetry $(\xi=0,\l=-1)$ can be written as
\beq \label{Qe}
       \mathcal{Q} = \int_{\bar{g}_{\m\n}} ^{g_{\m\n}}  \int_{\bar{A}_{\m}} ^{A_{\m}} \d \mathcal{Q}_{(0,-1)} \quad .
\eeq
The integration is intended between the background fields (which is Minkowski spacetime in this case, where the physical parameters $m,q,E$ are null) and the metric and gauge potential of the solution (\ref{metric})-(\ref{2.9}). The integrability property of the electric charge (\ref{Qe}) ensures that it is well defined, i.e. the result is independent on the parametric path followed for the integration.  \\
But, when dealing with the conserved energy, since we have no good criteria to define a canonical generator associated with the mass, the integrability of the charge can not be taken for granted. Therefore it would be more appropriate in (\ref{Qt}), as done by some authors in the literature, to use of a notation that emphasises this aspect, such as $\slashed{\delta} \mathcal{Q}_{(\p_t,0)}$. The difficulties in defining a canonical generator for the mass stem from the non-trivial asymtotic behaviour of the metric. In particular, in this static case, the main problem is the normalization of the Killing vector $\p_t$. As found in \cite{mass-mkn}, the act of imposing the integrability property on the mass charge $\mathcal{M}(m,q,E)$ uniquely selects its canonical generator. In the case under consideration this means that we can set the normalization of $\p_t$ by an integrating factor $\a(m,q,E)$ such that
\beq \label{dM}
           \d\mathcal{M} = \a \left[ \slashed{\d} \mathcal{Q}_{(\p_t,0)} - \Phi_{int} \d \mathcal{Q}  \right] \quad .
\eeq
The closeness of the one-form $\d \mathcal{\M}$ implies three partial differential equations in $(m,q,E)$. It is a non trivial fact that these three equations are fulfilled by just two function, the normalizing factor $\a(m,q,E)$ and the boundary electric chemical potential $\Phi_{int} (m,q,E)$, which remained undetermined. Here $\a(m,q,E)$ and $\Phi_{int} (m,q,E)$, which satisfy (\ref{alpha}) are
\bea \label{alpha}
       \a           &=&  \frac{4 (4+ q^2 E^2 )^{3/2}}{(4 - q^2 E^2)^2 \sqrt{4-4 \bar{A} E m q + E^2 q^2}} \quad , \\
       \Phi_{int}   &=& -A_{t_0}-\frac{E}{2 \bar{A}} + E \bar{A} q^2 \ \frac{4 - q^2 E^2}{4 + q^2 E^2} \quad .
\eea
Hence the integrability property is able to completely determine the canonical frame, setting a canonical time coordinate $t_{can}$ and furthermore setting the value of the electric chemical potential $\Phi_{int}$. Of course we can make use of the gauge freedom to select a frame where $\Phi_{int}=0$, by just choosing
\beq \label{At0}
 A_{t_0}=-\frac{E}{2 \bar{A}} + E \bar{A} q^2 \ \frac{4 - q^2 E^2}{4 + q^2 E^2} \quad .
\eeq
In this way $ A_{t_0} $ is uniquely fixed and, at the same time, this gauge choice makes the limit to the (non-accelerating) Reissner-Nordstrom solution well defined. \\
It can be verified that the surface where $\Phi_{int}=0$ lays inside the acceleration horizon, at least for a large range of the parameters $(m,q,E)$. \\
Having fixed the canonical frame through (\ref{alpha}) - (\ref{At0}), we can integrate the equation (\ref{dM}) to get the integrable mass for the accelerating black hole under consideration
\beq \label{Mass}
          \mathcal{M} =  \frac{ (4-q^2 E^2)^2 r_+^2 + 16 q^2 \bar{\D}_\varphi (1 -\bar{A}^2 r_+^2 ) }{2(4- q^2 E^2)^2 r_+ \sqrt{1-\bar{A}^2 r_+^2} } \sqrt{\bar{\D}_\varphi}
\eeq
We find remarkable that this expression for the mass can be cast in form of the Christodoulou-Ruffini  mass formula \cite{Christodoulou:1972kt}. In fact the mass (\ref{Mass}) can be expressed in terms of the event horizon area $\mathcal{A}$ (\ref{area}) and the intrinsic black hole electric charge $\mathcal{Q}$ (\ref{Q}) as follows:
\beq \label{mass-cr}
         \boxed{  \mathcal{M}^2 =  \frac{\mathcal{Q}^2}{2} + \pi \frac{\mathcal{Q}^4}{\mathcal{A}} + \frac{\mathcal{A}}{16 \pi} } \quad.
\eeq
Therefore the validity of the Christodoulou-Ruffini mass formula, even in the case where the black hole is drastically deformed by both the presence of acceleration and an external electromagnetic background, goes beyond its original derivation, for the asymptotically flat Kerr-Newman black hole. It suggests, also in view of the result of \cite{mass-mkn}, that this quantity encloses some fundamental properties of more general black holes. Actually in \cite{Ansorg:2010ru} this feature was intuited for axisymmetric extremal black holes, even when distorted by the presence of surrounding matter in Einstein-Maxwell theory.  \\
Considering that the entropy is a quarter of the area $\mathcal{S}:= \mathcal{A}/4$, and rescaling the Killing generator of the black hole horizon $\chi \rightarrow \bar{\chi}=\a \chi$, we can check that the first law of black hole thermodynamics is fulfilled, 
\beq
           \d \mathcal{M} = \bar{T} \d \mathcal{S} + \bar{\Phi} \d \mathcal{Q} \quad .
\eeq 
Note that the rescaling of $\chi$ induces a dilatation of the Hawking temperature $T$ and the Coulomb potential at the event horizon, according to the definitions (\ref{T})-(\ref{ks}) and (\ref{Phi}) respectively\footnote{Here we are considering the gauge choice (\ref{At0}) otherwise $\bar{\Phi}$ should be defined as $\a ( \Phi_{r_+} - \Phi_{int}$).} 
\bea \label{aT}
                \bar{T} &=& \frac{1}{2\pi} \sqrt{-\frac{1}{2} \nabla_\m \bar{\chi}_\n \nabla^\m \bar{\chi}^\n } \ \ \Bigg|_{r_+} =  \a T  \ \ , \\
     \label{aPhi}           \bar{\Phi} &=&  - \ \bar{\chi}^\m \ A_\m \ \bigg|_{r=r_+} \ = \qquad \a \Phi_{r_+} \ \ \quad .
\eea
Of course this rescaling naturally selects a canonical time, defined as $t_{can}= t/\a$.  With this coordinate transformation the $\a-$factor normalisation of the canonical Killing vector, which defines the event horizon, can be reabsorbed to read  $\bar{\chi}=\p_{t_{can}}$.\\
Moreover it can be straightforwardly checked that the canonical intensive quantities (constant everywhere on the event horizon) $\bar{T}$ and $\bar{\Phi}$ are coherent with the usual thermodynamic definitions
\bea
           \bar{T} &=& \frac{\p \mathcal{\M}}{\p  \mathcal{S}} =  \frac{1}{8 \pi \mathcal{M}} \left( 1 - \frac{\pi^2 \mathcal{Q}^4}{\mathcal{S}^2} \right) \quad , \\
           \bar{\Phi} &=&  \frac{\p \mathcal{\M}}{\p  \mathcal{Q}} =  \frac{\mathcal{Q}}{2 \mathcal{M S}} \left( \mathcal{S + \pi Q}^2 \right) \qquad .
\eea 
A further check of the relation between the mass $\mathcal{M}=\mathcal{\M(Q,S)}$, the extensive $(\mathcal{Q} , \mathcal{S})$ and the intensive $(\bar{T},\bar{\Phi})$ thermodynamic quantities computed above, is easily given by verifying the  Smarr formula
\beq \label{smarr}
         \mathcal{M} = 2 \bar{T} \mathcal{S} + \bar{\Phi} \mathcal{Q} \quad .
\eeq 
Note that the mass (\ref{Mass}) or (\ref{smarr}) is meaningful provided there is regularity of the spacetime, therefore the limit for null electric field $E$ have to be taken carefully. The acceleration constraint in the $E\rightarrow0$ limit reduces to  $\bar{A}_-=0$, in fact from a physical point of view, when there is no electric field the only possibility to have a regular spacetime is given by imposing null acceleration too. Hence taking the limit in the mass formula (\ref{Mass}) for vanishing acceleration and electric field (or, equivalently, commuting the limit order) we get the expected mass value $\mathcal{M}=m$, which coherently corresponds to the mass of the limiting spacetime, the  Reissner-Nordstrom black hole. \\
In case one is interested in the mass and thermodynamics of accelerating black holes, without the presence of a regularising electric or magnetic field background, the phase space method is not applicable anymore, precisely because of the conical singularity. However one might extrapolate the results we have found, for the regular case, in this section and in \cite{mass-mkn}, \cite{cft-duals-acc}.  In fact, assuming that, also in the non regular case, the mass fits into the Christodoulou-Ruffini formula, it is possible to fulfill the first law of black hole thermodynamics. Although with a different approach, the mass and thermodynamics for the accelerating Kerr-Newman black hole were presented in \cite{cft-duals-acc}. It is also possible to generalise these results in the presence of the cosmological constant. Of course the validity of the first law for irregular spacetimes should not to be taken for granted, therefore we will discuss this mass proposal in the Appendix \ref{app1}.\\

\section{Comments and Conclusions}

In this paper we have analysed the conserved charges and the thermodynamics of an accelerating Reissner-Nordstrom black hole, embedded in an external electric field background to ensure the regularity of the solution (outside the horizon).\\
While the electric charge is not problematic, to compute the mass we make use of the phase space technique which has the peculiarity to select a canonical time coordinate, and thus a canonical timelike Killing vector associated with energy by requiring the integrability of the conserved charge.\\
The resulting mass satisfies both the standard first law of thermodynamics and the Smarr formula. Moreover this mass fulfils the Christodoulou-Ruffini mass formula, originally derived for the Kerr-Newman black hole.  \\
Relying on this outcome we propose an analogous behaviour the in presence of the cosmological constant. In this case the solution presents irremovable conical singularities because it is not known how to embed these asymptotically (A)dS spacetimes in an external electromagnetic background, which is able to regularise the metric. Nevertheless, taking advantage of the Christodoulou-Ruffini formula for the Kerr-Newman-AdS black hole, it is possible to find a well-defined mass charge which fulfils the standard  first law of thermodynamics, without the need for extra ad-hoc constraints on the physical parameters, as it was proposed in previous literature.\\
More generically would be interesting to further investigate on the effectiveness of the Christodoulou-Ruffini mass formula, because it has also been shown to work well in the case of drastic deformation of the Kerr-Newman black hole. Perhaps it encloses some significant information about black holes, which remains undisclosed at the moment. Finally it would be worth exploring its range of applicability, also outside the realm of general relativity. \\

\section*{Acknowledgements}
{\small I would like to thank Geoffrey Compere, Roberto Oliveri and Cedric Troessaert for fruitful discussions. \\
This work has been funded by the Conicyt - PAI grant n$^\textrm{o}$ 79150061 and by Fondecyt project n$^\textrm{o}$ 11160945.} \\

\appendix

\section{Mass of the accelerating Kerr-Newman-AdS black hole}
\label{app1}

Accelerating (and rotating) black holes can be generalised in the presence of the cosmological constant\footnote{Which is better to think negative to avoid issues related to the presence of the cosmological horizon.} ($\L$). But, unfortunately, the Harrison transformation needed to embed these solutions in an external magnetic field does not work in this case, because some symmetries of the Einstein-Maxwell action are broken when $\L$ is not null \cite{ernst-lambda}. Therefore it is not possible to simultaneously eliminate both conical singularities from the rotating C-metric. This means that, physically, we still remain with at least one cosmological string (or strut).\\ 
Anyway, thanks to a negative cosmological constant, it is possible to select a parametric range where the inconvenient accelerating horizon \cite{chen-teo} is absent, obtaining the so-called ``slowly accelerating" black holes. \\
The rotating and accelerating C-metric that we consider, in spherical coordinates ($ x:= \cos \theta $), is  
\begin{align} \label{rot-c+l}
                  ds^2 = \frac{1}{ (1 + A x r)^2 }   & \left[  \frac{f(r) + a^2 h (x)}{r^2 + a^2 x^2 } dt^2  - \frac{r^2 + a^2 x^2 }{f(r)} dr^2         +  \frac{r^2 + a^2 x^2 }{h(x)} dx^2 \right. \\
                 &   \left.  + \frac{a^2 \D_\varphi^2 (1-x^2)^2 f(r) + (a^2+r^2)^2 h(x)}{r^2+a^2 x^2}   d\varphi^2  + 2 \frac{a (1-x^2) f(r) + a (a^2+r^2) h(x) }{r^2+a^2 x^2} \D_\varphi dt  d\varphi  \right] \  , \nn
\end{align}
where 
\bea
        f(r) &:=& (A^2 r^2 -1) (r-r_+) (r-r_-) + \frac{\Lambda}{3} \left( r^4 + \frac{a^2}{A^2} \right) \ \ , \\
        h(x) &:=& (1 - x^2) (1+ A x r_+) (1+ A x r_-) \quad .
\eea
Its associated electromagnetic field is the same as the Kerr-Newman solution
\beq
       A_\m = \left[- \frac{q r + p a x}{r^2 + a^2 x^2} , 0 , 0 , - \frac{ - q r a ( 1-x^2 ) - p x ( r^2 + a^2 ) }{r^2 + a^2 x^2}    \right]	  \quad .
\eeq
In fact the (A)dS-Kerr-Newman metric can be recovered in the vanishing accelerating limit\footnote{Just a little care in rescaling the time and radial coordinate is needed before taking the $A\rightarrow0$ limit.}. 
Note that the in presence of the cosmological constant the positions of the inner ($r_i$) and event horizon ($r_h$) is no longer located at $r_-$ and $r_+$; rather they are more algebraically involved. They correspond to the real roots of the quartic equation $f(r)=0$. In \cite{hong-teo} the following parametrization it is proposed: 
\beq \label{rpmpara}
          r_\pm = m \pm \sqrt{m^2-q^2-p^2 -a^2\left( 1+ \frac{\L}{2 A^2} \right)}   \quad  .
\eeq 
However, in order to fulfil the Einstein equation of motion, it is sufficient only a constraint on $r_-$, as follows 
\beq
                r_- = \frac{3 A^2 (q^2+p^2) + a^2 (3 A^2 + \L)}{3 A^2 r_+} \quad  .
\eeq
To remove the conical singularity, as done in Section \ref{acc-ectr}, from the north pole we have to set the constant $\D_\varphi $ to a specific value
\beq
          \bar{\D}_\varphi = \frac{1}{(1 +  A r_+ ) (1 + A r_-)} \quad .
\eeq  
Alternatively, it is possible to fix $\D_\varphi $ to remove the conical singularity from the south pole.
Henceforward the magnetic charge $p$ will be considered null to ensure regularity of the electromagnetic field on the axis of symmetry, i.e. to avoid Dirac strings. \\
The event horizon area 
\beq \label{Acony}
     \mathcal{A} = \int_0^{2\pi} d\varphi \int_{-1}^1 dx \sqrt{g_{\varphi \varphi} g_{xx}} \ = \   \frac{ 4 \pi \D_\varphi  (r_h^2 + a^2)}{1-A^2 r_h^2} \quad ,
\eeq
is, as usual, related to the black hole entropy by 
\beq
           \mathcal{S}=\frac{\mathcal{A}}{4} \quad .
\eeq
The electric charge and angular momentum are respectively given by
\bea  \label{Qcony}
           \mathcal{Q} &=& \frac{1}{8\pi} \int_\S F^{\m\n} d\S_{\m\n} =   q \ \D_\varphi \quad ,  \\
 \label{Jcony} J &=& \frac{1}{16\pi}   \int_ {\S} \left[ \nabla^\a \xi^\b_{(\varphi)} + 2 F^{\a\b} \xi^\m_{(\varphi)} A_\m \right] d\S_{\a\b}   = a m \D_\varphi^2 \quad ,
\eea
where $\xi^\m_{(\varphi)}$ represents the rotational Killing vector $\p_{\varphi}$.  \\
In the chosen gauge (\ref{rot-c+l}) the Coulomb electric potential $\Phi_e$ and the angular velocity at the horizon $\Om_h$ are 
\bea \label{Phie}
      \Phi_e &:=& - \chi^\m A_\m \Big|_{r=r_h} \ =  \frac{q \ r_h}{r_h^2 +a^2} \quad ,\\
 \label{Omh}  \Om_h &:=& -\frac{g_{t \varphi}}{g_{\varphi \varphi}} \ \ \Bigg|_{r=r_h} \ = - \frac{a}{(r_h^2+a^2) \D_\varphi} \quad .
\eea
Defining the Killing generator of the horizon as $\chi = \p_t + \Om_h \p_\varphi$, we get the temperature of the black hole event horizon from the surface gravity $\k_s$ 
\beq
          T = \frac{\k_s}{2\pi} := \frac{1}{2\pi} \sqrt{-\frac{1}{2} \nabla_\m \chi_\n \nabla^\m \chi^\n } \ \ \Bigg|_{r_h} = 
            \frac{-3 a^2 - 2 q^2 + 3 m (r_h + A^2 r_h^3) + r_h^4 (3 A^2 + \L) }{6 \pi r_h (a^2+r_h^2)} \quad .
\eeq
In Section \ref{acc-ectr} and in \cite{mass-mkn} we have found that, also in the presence of atypical asymptotics such as the accelerating and the magnetised ones, when the black hole is regular, the mass obey to the Christodoulou-Ruffini mass formula \cite{Christodoulou:1972kt}. Assuming that this behaviour also holds in the presence of a conical singularity we can write a mass formula for the accelerating (A)dS-black hole, which obeys the standard first law of black hole thermodynamics\footnote{For some values of  $\a , \Om_{int}$ and $\Phi_{int}$, as it is explained in \cite{mass-mkn}. $\Om_{int}$ and $\Phi_{int}$ depend on the gauge choice, they can be null for a suitable observer.} 
\beq \label{1law}
       \d\mathcal{M} = \bar{T} \d \mathcal{S} + \bar{\Om} \d \mathcal{J} + \bar{\Phi} \d\mathcal{Q} \quad ,
\eeq
where, as done in \cite{mass-mkn} and coherently with  Section \ref{mass}, $\bar{T} , \bar{\Omega}$ and  $\bar{\Phi}$ are defined as follows
\beq \label{barall}
      \bar{T} = \a T \quad , \qquad \bar{\Om} = \a (\Omega_h - \Om_{int}) \quad , \qquad \Phi = \a (\Phi_h - \Phi_{int}) \quad . 
\eeq
Actually, in absence of the cosmological constant, we have already verified in \cite{cft-duals-acc} that, when the first law is imposed to find the mass, it gives an expression which exactly coincides with the Christodoulou-Ruffini mass formula. But that path is technically more difficult to pursue when the cosmological constant is present. Nevertheless the generalisation of the Christodoulou-Ruffini mass formula is available \cite{caldarelli} and it is given by
\beq \label{masscony}
        \mathcal{M}^{2}(\mathcal{S}, \mathcal{J}, \mathcal{Q})=\frac{\mathcal{S}}{4\pi} + \frac{\mathcal{Q}^{2}}{2} - \frac{\L}{3} \mathcal{J}^2 +\frac{\pi (\mathcal{Q}^4+4\mathcal{J}^2)}{4\mathcal{S}} - \frac{\L}{3} \ \frac{\mathcal{S}}{2\pi} \left( \mathcal{Q}^2  + \frac{\mathcal{S}}{\pi}  - \frac{\L}{3} \frac{\mathcal{S}^2}{2\pi^2}    \right) \ \ .
\eeq
Finally, substituting in (\ref{masscony}) the values of the extensive quantities $\mathcal{S}, \mathcal{Q}, \mathcal{J},$ (\ref{Acony}) - (\ref{Jcony}), we get the  mass for the accelerating AdS-Kerr-Newman black hole, in terms of the metric parameters ($m,a,q,A$). \\
Note that $\a , \Om_{int}$, and $\Phi_{int}$ can be uniquely inferred by the thermodynamic intensive parameters conjugate to $\mathcal{S,J}$ and $\mathcal{Q}$, as found in \cite{caldarelli}
\begin{align} \label{Tbar}
                  \bar{T}   = & \ \frac{\p \mathcal{M}}{ \p \mathcal{S}} \bigg|_{\mathcal{JQ}} \ = \ \frac{1}{8\pi \mathcal{M}} \left[ 1 - \frac{\pi^2}{\mathcal{S}^2} \left( 4 \mathcal{J}^2 + \mathcal{Q}^4 \right) - \frac{2 \L}{3} \left( \mathcal{Q}^2 + \frac{2 \mathcal{S}}{\pi} \right)  + \frac{\L^2}{3} \frac{\mathcal{S}^2}{\pi^2} \right] \quad ,  \\
              \label{Ombar}    \bar{\Om} = &  \ \frac{\p \mathcal{M}}{ \p \mathcal{J}} \bigg|_{\mathcal{SQ}}  \  = \ \frac{\pi \mathcal{J}}{\mathcal{M} \mathcal{S}} \left( 1-\frac{\L \ \mathcal{S}}{3 \ \pi} \right) \quad  , \\
               \label{Phibar}   \bar{\Phi} = & \ \frac{\p \mathcal{M}}{ \p \mathcal{Q}} \bigg|_{\mathcal{SJ}} \ = \frac{\pi \mathcal{Q}}{2 \mathcal{M} \mathcal{S}} \left( Q^2 + \frac{\mathcal{S}}{\pi	} -\frac{\L}{3} \frac{\mathcal{S}^2}{\pi^2}  \right)    \quad .      
\end{align}
Note the conceptual difference with respect to the case without the cosmological constant, as in Section \ref{mass} or in \cite{mass-mkn}, where, since the metric was regular it was possible to directly compute $\d\mathcal{M}$ and derive $\a$, $\Phi_{int}$ (and $\Om_{int}$) from the mass integrability requirement. Only afterwards was the agreement with the thermodynamic quantities verified.  But in this case, where the metric is irregular due to the presence of the cosmological constant, $\a , \Om_{int}$ and $\Phi_{int}$ can be determined by assuming  the validity of the thermodynamic relations (\ref{Tbar})-(\ref{Phibar}). \\

Instead of interpreting the mass, from a thermodynamic point of view, as the energy of the black hole, it can be interpreted as enthalpy \cite{sourya}, \cite{dolan}. In this case the cosmological constant is not considered fixed anymore, but it is allowed to fluctuate. From the cosmological constant it is possible to define a pressure term, $\mathcal{P}=-\frac{\L}{8\pi}$, which is treated as thermodynamic state variable, while the thermodynamic variable conjugated to $\mathcal{P}$ is given by the volume of the black hole
\beq
                 V :=   \frac{\p \mathcal{\M}}{\p  \mathcal{P}} \bigg|_{(\mathcal{S},\mathcal{J},\mathcal{Q})} 
                    =    \frac{1}{\mathcal{M}}    \left[ \frac{4\pi}{3} \mathcal{J}^2 + \frac{2 \mathcal{S}}{3} \left( \mathcal{Q}^2 + \frac{\mathcal{S}}{\pi} \right) + \frac{16}{9} \mathcal{P} \mathcal{S}^3 \right]                 \quad .
\eeq
In this settings an extra term in the first law of black hole thermodynamics has to be included as follows:
\beq \label{new1law}
          \d\mathcal{M} = \bar{T} \d \mathcal{S} + V \d \mathcal{P} + \bar{\Om} \d \mathcal{J} + \bar{\Phi} \d\mathcal{Q} \quad .
\eeq
The new term $ V \d \mathcal{P}$ refers to the change of enthalpy of the system for a pressure fluctuation $\d\mathcal{P}$, at constant volume.\\
Remarkably the Christodoulou-Ruffini mass formula (\ref{masscony}) satisfies not only the first laws (\ref{1law}) and (\ref{new1law}), but also the Smarr formula
\begin{align}
       \mathcal{M} & = 2 \bar{T} \mathcal{S} - 2 V \mathcal{P} + 2 \bar{\Om} \mathcal{J} + \bar{\Phi} \mathcal{Q} \\
                   & = 2 \bar{T} \mathcal{S} + 2 \bar{\Om} \mathcal{J} + \bar{\Phi} \mathcal{Q} - \frac{4 \mathcal{S P}}{3} \left[ \frac{\bar{\Phi} \mathcal{S}}{\pi \mathcal{Q}} + \frac{\bar{\Om} \mathcal{J}}{1+8 \mathcal{PS}/3}\right]  \quad .
\end{align}
We stress that in this approach no ad-hoc assumptions, boundary conditions or constraints on the parametric space are needed to satisfy the first laws and the Smarr formula. Anyway it strongly relies on the assumption of the validity of the Christodoulou-Ruffini mass formula for the accelerating and singular C-metrics we are considering in this Appendix.\footnote{After the publication of this paper another approach to study black hole thermodynamics in presence of conical singularities is carried out in \cite{ruth-cony}. It is based on the modification of the structure of the first law due to the presence of the nodal singularities.} \\
To give an example we will explicitly work out the non rotating case, i.e. $a=0$ and $\mathcal{J}=0$. Thus we are considering an accelerating Reissner-Nordstrom-AdS black hole.\\
From (\ref{Tbar}) and (\ref{Phibar}) we find, making use of  (\ref{Phie}), (\ref{Omh}) and (\ref{barall}), $\a$ and $\Phi_{int}$ respetively
\begin{align}
                   \a \ = \ & \ \frac{3 r_h^4 \D_\varphi \L - 3 (1-A^2 r_h^2) \left[ r_h^2 - q^2 (1-A^2 r_h^2) \D_\varphi  \right] }{2 \sqrt{1-A^2 r_h^2} \sqrt{\D_\varphi} \left[ 3 q^2 -3 m (r_h+A^2 r_h^3) + r_h^4 (3A^2 + \L) \right] } \quad , \\
                  \Phi_{int}  \ = \ & \frac{q}{\a r_h} \left( \a - \sqrt{\D_\varphi} \sqrt{1-A^2 r_h^2} \right) \quad .
\end{align}
These quantities exactly reduce to the ones found in \cite{cft-duals-acc} in the case of a null cosmological constant. Finally, the mass, using the parametrization (\ref{rpmpara}) for $r_\pm$, reads
\beq \label{massa0lambda}
         \mathcal{M}\Big|_{a=0} \ = \ m \ \sqrt{\frac{1+Ar_h }{(1-Ar_h)(1+2mA+q^2A^2)^3}}
\eeq
This value for the mass is in coherence with the special case treated in \cite{cft-duals-acc}, and in the null accelerating limit it goes to $m$, the mass of the Reissner-Nordstrom-AdS black hole, as expected. However, the mass (\ref{massa0lambda}) does not coincide with other values in the literature, as in \cite{ruth}. Therefore the thermodynamic analysis is also different. The main difference lies in the normalisation of the timelike Killing vector, which we choose, according to our previous results based on the covariant phase space methods \cite{mass-mkn} and Section \ref{mass}, in agreement with the Christodoulou-Ruffini mass formula. In our opinion this proposal seems more well grounded than the previous, because, without extra ad hoc assumptions on the black hole physical parameters, the first law of black hole thermodynamics both in the standard case, where the mass is considered as the internal energy of the thermodynamic system, and in the case where the mass is associated with the enthalpy are fulfilled. Moreover it is coherent with the phase space techniques, when the cosmological constant vanishes. \\

\end{document}